# Scaling of dissipation in megahertz-range micromechanical diamond oscillators

Matthias Imboden and Pritiraj Mohanty [a)]
*Department of Physics, Boston University, 590 Commonwealth, Boston, Massachusetts 02215*
Alexei Gaidarzhy
*Department of Aerospace and Mechanical Engineering, Boston University, 110 Cummington Street, Boston, Massachusetts 02215*
Janet Rankin and Brian W. Sheldon
*Division of Engineering, Brown University, Providence, Rhode Island 02912*

**We report frequency and dissipation scaling laws for doubly-clamped diamond resonators. The device lengths range from *10 μm* to *19 μm* corresponding to frequency and quality-factor ranges of *17 MHz* to *66 MHz* and 600 to 2400 respectively. We find that the resonance frequency scales as *1/L²* confirming the validity of the thin-beam approximation. The dominant dissipation comes from two sources; for the shorter beams, clamping loss is the dominant dissipation mechanism; while for the longer beams, surface losses provide a significant source of dissipation. We compare and contrast these mechanisms with other dissipation mechanisms to describe the data.**

Understanding relevant dissipation mechanisms is important for the development and design of high-frequency micromechanical and nanomechanical resonator-based devices. In these structures, dissipation is usually the limiting factor for performance and sensitivity in sensing and signal processing[1,2,3]. The central problem lies in identifying and minimizing the dominant mechanisms of dissipation (inverse of the quality factor, Q).

Even though the scaling of frequency with system size on the micron scales is now being routinely demonstrated, the corresponding scaling of dissipation is yet to be comprehensively studied. Specifically, dissipation mechanisms such as clamping loss, thermoelasticity and anharmonic mode-coupling can be compared and contrasted with the contributions arising from internal defects in the bulk and the surface. Usually, loss mechanisms due to the coupling of resonance modes to substitutional and configurational defects are better understood at low temperatures by the tunneling[4,5] and activation effects[6]. The geometry-dependent effects such as clamping loss, however, can be explored in relatively high temperature ranges. Here, we report a comprehensive set of measurements of resonance frequency and dissipation in micromechanical structures of diamond. For the resonance frequency, we observe a scaling dependence of $(t/L^2)$, expected from the thin-beam approximation in continuum elasticity. We find that the corresponding scaling of dissipation enables the identification of the dominant dissipation mechanisms.

Diamond is the ideal material for micro- and nano-mechanical systems. It has the highest known acoustic velocity $\sqrt{E/\rho}$ and superior thermal properties, important for manufacturing high frequency structures. Diamond is also chemically inert, making it an attractive material for MEMS and NEMS fabrication. Here, we present scaling laws in micromechanical oscillators, to better understand what dissipation mechanisms contribute to loss in diamond doubly-clamped oscillators. All data is taken in the high temperature range, between 230 and 290K.

The nanocrystalline diamond (NCD) films used to fabricate the beams were grown by microwave plasma CVD (AsTex HPM/M system) with a gas mixture of 1% $CH_4$, 5% $H_2$, and 94% Ar. More detailed descriptions of NCD produced with this method are provided elsewhere[7]. The films for the current work were deposited on oxidized Si substrates at 800 C for 100 min, with a microwave power of 800 W. It is generally more difficult to initiate NCD growth on these silica surface (compared to Si), thus careful attention to the seeding process was necessary. Before deposition, the samples were ultrasonically cleaned using sequential baths of acetone and methanol. Samples were subsequently submerged in a dispersion of nanocrystalline diamond powder and methanol, ultrasonicated for 35 minutes, and then rinsed thoroughly with methanol prior to deposition. Sample set #2 was obtained from a substrate that was positioned underneath the center of the plasma during growth, while sample set #1 was obtained from a substrate that was approximately 2 cm away from the center of the plasma.

We fabricate the micromechanical oscillators using standard e-beam lithography and surface nanomachining. The 50-nm-thick gold electrodes are evaporated thermally onto the diamond, a thin layer of titanium aids adhesion. A 600-nm chromium layer evaporated on top provides a mask to withstand the RIE. The diamond is etched anisotropically in an $O_2$ (50sccm)/$CF_4$ (2sccm) RIE plasma at a pressure of 30 mTorr. At 300W the etch rate is approximately 50 nm min$^{-1}$. Buffered HF isotropic wet etch removes the silicon oxide beneath the diamond beams. Figure 1 displays a series of suspended diamond structures fabricated by this method. Each sample set contains three harp structures and a readout electrode; each harp is made up of four mechanically isolated oscillators. The oscillator lengths $L$ vary from *8 μm* to *19 μm* for sample set #1 and *6 μm* to *17 μm* for sample set #2 in one micron steps. The width $w$ is *500 nm* by design and the thickness $t$ measured using an SEM is *1100 nm* and *800 nm* for set #1 and #2 respectively.

The oscillators were actuated and the response recorded using the magnetomotive method[8]. We placed the samples under vacuum (< $10^{-5}$ Torr) in a homogenous 8T magnetic field; we measured set#1 at temperatures ranging from 230 K to 240K and set#2 at 280K to 290K. Both temperature ranges are significantly above the Debye peak[9]. An HP network analyzer provided for actuation and readout. A pre-amp was used to enhance the signal. The oscillators were all excited in the in-plane mode (the displacement is in parallel to the diamond film). The balanced-bridge technique[10] further enhances the signal by nulling the background. Schematics of the setup are shown in figure 1 (a).

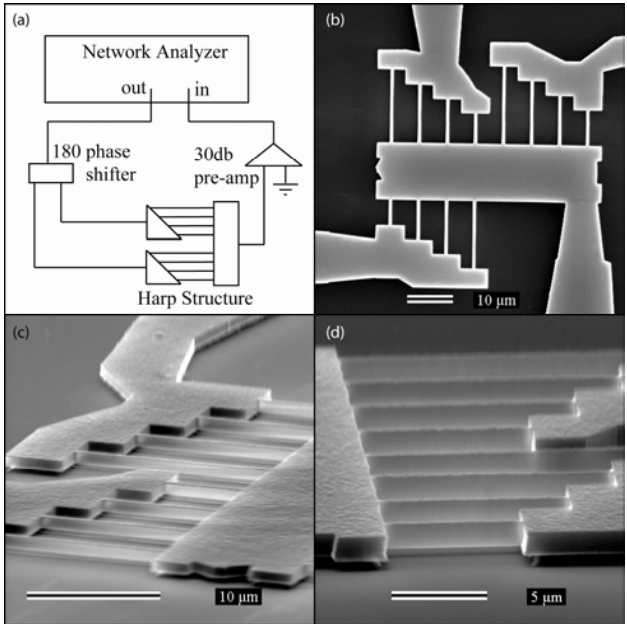

**Figure 1.** Measurement electronics setup and SEM pictures of diamond harp samples. (a) Actuation and readout circuit with HP network analyzer. The balanced bridge technique reduces background noise. (b) and (c) Top and tilted view of sample set *#2*. The set includes three harps each with four doubly clamped beams. Beam lengths $L$ vary from *6 μm* to *17 μm*, the width $w$ = *500 nm* and thickness $t$ = *800 nm*. (d) Tilted SEM picture of sample set #1. $L$ varies from *8 μm* to *19 μm*, $w$ = *500 nm* and $t$ = *1100 nm*.

A total of 17 resonances were found, 10 for sample set #1 and 7 for set #2, some of which are presented in figure 2 (a). The symmetric shapes of the lorentzians indicate that the response is linear for the applied drive power. Plotting the resonance frequency vs. $L^{-2}$ shows the linear relationship as predicted by the thin-beam approximation[11] (see figure 2 (b)). The slope derived from a fit to $f_0 = 1.03 \sqrt{E/\rho}\, t/L^2$ can be used to measure $\sqrt{E/\rho}$ directly and hence the Young's modulus $E$. We find the Young's modulus of sample set #1 to be $E$ = *678±62 GPa* and for set #2 $E$ = *981±89 GPa*. The higher modulus is very close to values obtained with nanoindentation and acoustic wave measurements, for films grown under similar conditions. The lower modulus value for sample set #1 is consistent with the lower plasma density during the growth of this material. Based on more detailed characterization of similar films, this material probably contains more $sp^2$ carbon, most of which is believed to be associated with bonding at grain boundaries[12]. For both samples, the Young's modulus is much greater than that of silicon.

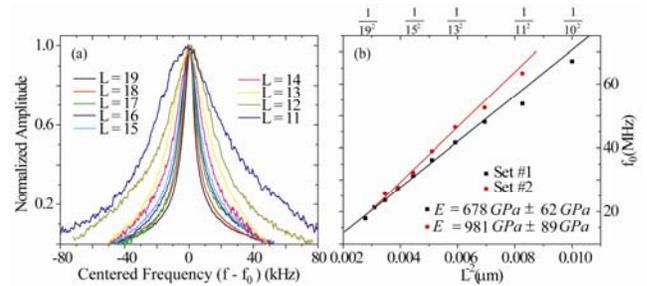

**Figure 2.** (Color online) (a) Lowest nine resonances found for sample set #1 centered on zero and normalized amplitude. Resonance frequencies range from $f_0$ = *17.908 MHz* for $L$ = *19 μm* to $f_0$ = *67.140 MHz* for $L$ = *10 μm*. (b) Frequency scaling for both sample sets. The thin beam approximation predicts $f_0 = 1.03 \sqrt{E/\rho}\, t/L^2$, and provides a very good fit to observations. From the slope of the linear fit, we calculated the Young's modulus for both samples.

Figure 3 depicts the dissipation $Q^{-1}$, which is the inverse of the quality factor ($Q = f_0/\Delta f$) for both sample sets calculated from the Lorentzian fits to the resonances. Over this rather short range in length scale, it is difficult to determine power laws with high confidence. For large $L$, sample set#2 shows somewhat lower dissipation. The dashed lines are fits indicating the scaling laws. For both figures, the effects of the magnetic field on the damping are subtracted out using the data from the magnetic field dependent dissipation plot (see insert of figure 3). In the linear regime, the dissipation is drive power independent. Magnetomotive damping given by $Q^{-1} = (\xi L^2 B^2 R)/(2\pi m f_0 |Z|^2)$ (Ref. [8]) is an artifact caused by the actuation and readout method and is not related to the material used, note that there is also the possibility if intrinsic magnetic field dependence on dissipation, that has nothing to do with the readout circuit[6]. We observe good fits for both $Q^{-1} = aL^{-3} + b$ and $Q^{-1} = aL^{-5} + b$, where $b$ parameterizes the length independent dissipation, electrode losses are believed to be negligible[14].

Dissipation in oscillators can come from many different sources. Thermoelastic losses, prevalent in high temperature measurements for large beams, are not detectable in these structures. The dissipation shows a significant dependence with magnetic field, especially for the longer low-frequency beams. The origin of this behavior is the sum of magnetomotive damping and intrinsic sources. This source of loss can be either calculated or determined experimentally, as it is done here (see insert to figure 3), and subtracted out, where we assumed a $Q^{-1} \sim a\frac{L^2}{f_0}B^2$ dependence and $a$ is extracted from the insert in figure 3. We find that clamping losses are probably the most significant source of dissipation in the shorter beams; and the dissipation becomes length independent for the longest beams. In this parameter space dominated by surface losses, the contributions to dissipation in the diamond itself become significant. We fit the data to relevant models to determine the best agreement. The results are summarized in Table I.

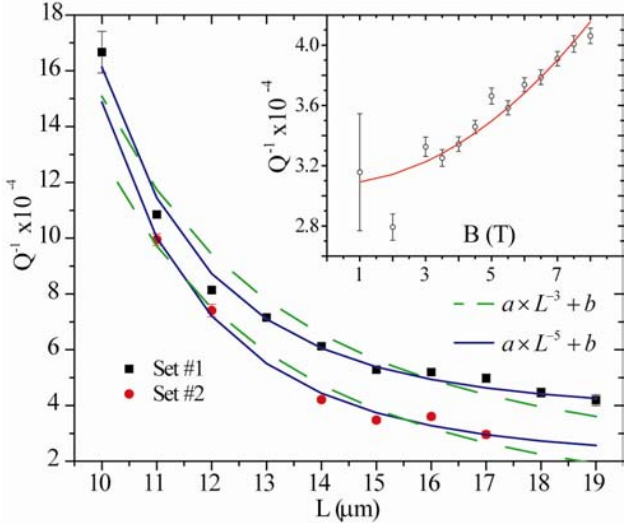

**Figure 3.** (Color online) Dissipation ($Q^{-1}$) plots for both sample sets. Two fits are illustrated $aL^{-n} + b$ with $n=\{3,5\}$. For both sets, $n=5$ appears to be the best fit for the scaling law; due to the small range in length-scale, one can not exclude $n=3$ either. The constant $b$ portrays the length independent dissipation attributed mostly to surface losses. The magnetic field dependence of dissipation, including a quadratic fit, is illustrated in the inset.

Thermoelastic damping has been calculated using standard models[13]. For structures smaller than 30 μm and high frequencies, thermoelastic damping is not the dominant contribution. The theoretical numbers $Q^{-1}_{TED}$ for our sample are on the order of $10^{-12}$, many orders of magnitude smaller than the observed values. The effects of magnetomotive damping are found to be significant for the longer low-frequency beams. This effect is subtracted out of the dissipation during the data analysis, as discussed above, and illustrated in the insert of figure 3.

From the fit of the dissipation dependence to beam length, we can compare different clamping loss theories to our findings. Hao, Erbil and Ayazi[15] model two-dimensional doubly-clamped beam structures and find $Q^{-1}_{clam\ in\ plane} = \alpha(w/L)^3$, where we determine experimentally $\alpha=10.7$ and $\alpha=10.3$ for set #1 and set #2 respectively. This is in good agreement with previous measurements of diamond oscillators[9]. The $L^{-3}$ dependence fits reasonably well to the data, where the rather small range in beam lengths makes it difficult to exclude other possibilities. For the out-of-plane model the dissipation is predicted to be $Q^{-1}_{clam\ out\ of\ plane} = \beta(w/L)$. Not surprisingly, this is a poor fit to observations. On the other hand strong length dependencies, such as the $L^{-5}$ fit appear to be a good fit to observation. The two-dimensional dissipation model does not consider the influence of the oscillator thickness or support structure. An extended support structure for out-of-plane actuation has shown to result in a strong length dependency for the dissipation[16]. Possibly more involved theories need to be developed that include such contributions, for the in-plane actuation. Consideration should include not only the extended base structure but also the effects of the isotropic wet etch of the sacrificial silicon oxide and the multiple materials involved.

| Theory | $Q^{-1}$ | Set#1 | Set#2 |
| --- | --- | --- | --- |
| magnetomotive damping | $\dfrac{\xi L^2 B^2 R}{2\pi m f_0 |Z|^2}$ | 0.000026 - 0.00018 | 0.000013 - 0.00018 |
| clamping in plane | $\alpha(w/L)^3$ | $\alpha = 10.7$ | $\alpha = 10.3$ |
| Clamping out of plane | $\beta w/L$ | $\beta = 0.045$ | $\beta = 0.043$ |
| surface losses | $\dfrac{\gamma 2\delta(3w+t)}{wt}\dfrac{E_{ds}}{E}$ | $\gamma\delta E_{ds}=11.5$ | $\gamma\delta E_{ds}=0.29$ |
| Electrode losses | $\dfrac{Q_{sp3}^{-1} + \beta Q_{Au}^{-1}}{1+\beta}$ | <1% | <1% |

**Table I**. Fitted parameters to various dissipation models. $\alpha$ is in good agreement with similar experiments (see Ref. 9). The in-plane clamping losses model is a good fit to observations. For other scaling laws, there is poor numeric agreement with similar studies (see Ref. 17). Surface losses may explain the major source of dissipation for the $L$ independent regime. Loss from the electrodes (see Ref. 14) can be neglected, since $\beta=t_{Au}E_{Au}/t_{sp3}E_{sp3}<<1$.

The measured dissipation illustrates that a significant loss component originates from length independent contributions. A good candidate for this mechanism is surface loss (a rather general term including for example contributions from surface defects, impurity atoms, dangling bonds, and crystal defects), which can be modeled as $Q_{SL}^{-1} = \frac{\gamma 2\delta(3w+t)}{wt}\frac{E_{ds}}{E}$ (Ref. [17]). Generally, it will contribute when the surface to volume ratio is large or if there are inherent surface stresses, which are known to occur during CVD diamond growth. The parameters are difficult to determine theoretically. Using the fit $\alpha\, L^{-3} + \beta$, we find $\gamma\delta E_{ds}=11.5$ and $\gamma\delta E_{ds}=0.29$ for sample set #1 and #2 respectively. These 2 distinct values may be explained by differences in surface stress, surface roughness, or surface chemistry originating during CVD growth. This observed variation of more than a factor of 30 suggests that the growth process can have a substantial impact on the surface properties and losses in these materials.

In conclusion, the dissipation mechanisms in these diamond RF oscillators at high temperatures can be well explained by clamping losses and surface losses only. We find that the two-dimensional model for clamping losses for in-plane excitations reproduces the observed dependence in the data. Both qualitative and quantitative agreements of the data with analytical expressions for dissipation are useful in the development of mechanical oscillators and resonator-based sensing and actuating devices in the RF-frequency range.

Dr. Xingcheng Xiao's assistance with the development of the seeding and growth procedures is gratefully acknowledged. Diamond fabrication at Brown University is supported by NSF (DMR-0305418). The work at Boston University is supported by NSF (DMR-0449670 and CCF-0432089).

[a] E-mail: mohanty@buphy.bu.edu